\documentclass[reprint,twocolumn,superscriptaddress,prl,showpacs]{revtex4-1}
\usepackage{graphicx}
\usepackage{dcolumn}
\usepackage{bm}
\usepackage{amsmath}
\usepackage{natbib}

\usepackage[T1]{fontenc}
\usepackage[letterpaper,textwidth=7in,top=.75in,bottom=.75in]{geometry}
\begin{document}
\title{Temperature and magnetic field dependent longitudinal spin relaxation in nitrogen-vacancy ensembles in diamond}

\author{A.~Jarmola}
 \email{jarmola@berkeley.edu}
    \affiliation{
     Department of Physics, University of California,
     Berkeley, California 94720, USA
     }
    \affiliation{
    Laser Centre, The University of Latvia, Rainis Boulevard 19, 1586 Riga, Latvia
    }
 \author{V.~M.~Acosta}
    \affiliation{
     Department of Physics, University of California,
     Berkeley, California 94720, USA
    }
     \affiliation{
     Hewlett-Packard Laboratories, 1501 Page Mill Road, Palo Alto, California 94304, USA
    }

 \author{K.~Jensen}
    \affiliation{
     Department of Physics, University of California,
     Berkeley, California 94720, USA
    }

\author{S.~Chemerisov}
    \affiliation{
     Argonne National Laboratory, Argonne, Illinois 60439, USA
    }

 \author{D.~Budker}
 \email{budker@berkeley.edu}
    \affiliation{
     Department of Physics, University of California,
     Berkeley, California 94720, USA
    }
    \affiliation{Nuclear Science Division, Lawrence Berkeley National Laboratory,
     Berkeley, California 94720, USA
    }

\date{\today}

\begin{abstract}
We present an experimental study of the longitudinal electron-spin relaxation time ($T_{1}$) of negatively charged nitrogen-vacancy (NV) ensembles in diamond. $T_{1}$ was studied as a function of temperature from 5 to 475 K  and magnetic field  from 0 to 630 G for several samples with various NV and nitrogen concentrations. Our studies reveal three processes responsible for $T_{1}$ relaxation. Above room temperature, a two-phonon Raman process dominates, and below, we observe an Orbach-type process with an activation energy, 73(4) meV, which closely matches the local vibrational modes of the NV center. At yet lower temperatures, sample dependent cross relaxation processes dominate, resulting in temperature independent values of $T_{1}$, from ms to minutes. The value of $T_{1}$ in this limit depends sensitively on magnetic field and can be tuned by more than an order of magnitude.

\end{abstract}


\maketitle

Negatively charged nitrogen-vacancy (NV) color centers in diamond have been a focus of recent research due to their promise as fluorescent markers for biological systems \cite{FU2007, FAK2009, MCG2011}, qubits that can be optically initialized and readout \cite{NEU2010, TOG2010, ROB2011NAT}, and sensors of magnetic \cite{TAY2008, BAL2008, MAZ2008NATURE, ACO2010APL} and electric \cite{DOL2011} fields. The NV center is uniquely suited for these applications due to its atom-scale spatial resolution and exceptional optical and spin properties over a wide range of operating temperatures.

 The spin phase coherence time ($T_{2}$) is a critical figure of merit for these emerging quantum-based applications. For example, for ensemble magnetometry, the sensitivity scales as $(N T_{2})^{-1/2}$, where $N$ is the number of spins. Therefore, optimization of ensemble sensors \cite{ACO2009, ACO2010APL, MAE2010, PHA2011} calls for high-density samples with long phase coherence times. In quantum computing, $T_{2}$ constrains the minimum gate operation time and limits the performance of quantum error correction protocols \cite{DIV2000, LAD2010}. Typically $T_{2}$ is limited by magnetic impurities in the local environment which serve as sources of decoherence \cite{ABR1970}. However, it is possible to reduce the effect of the spin bath using dynamic-decoupling techniques \cite{DEL2010SCIENCE, RYA2010, NAY2011, PHA2012}. The phase coherence time is then limited by stochastic processes, such as phonon interactions, which cause irreversible changes in axial spin projection. Such processes are known as longitudinal relaxation, with a characteristic timescale, $T_{1}$.

The temperature dependence of $T_{1}$ has been experimentally studied previously at high magnetic fields using electron paramagnetic resonance \cite{RED1991, TAK2008, HAR2006}. In Ref. \cite{RED1991}, $T_{1}$ was measured in the range 100-500 K, and the results were described by a model including Orbach and Raman phonon processes. Other results \cite{TAK2008} over a somewhat lower temperature range could not be accurately described by this model. In Ref. \cite{HAR2006}, $T_{1}$ was measured at T $\approx$ 2 K for two different samples, and the results differed by more than an order of magnitude, casting further doubt that $T_{1}$ can be explained purely by phonon processes.

In this Letter, we report an investigation of longitudinal spin relaxation $T_{1}$ of NV ensembles as a function of sample impurity content, temperature, and magnetic field. We identify the principal relaxation mechanisms as interactions with local and lattice phonons, as well as cross relaxation with nearby spin impurities. We develop a universal model for $T_{1}$ relaxation of NV centers which agrees well with experiments on samples across a wide range of impurity concentrations and temperatures.

\begin{table}[ht]
\caption{\label{tab:values} Sample characteristics (see Ref. \cite{ACO2009} for details). The diamonds were irradiated with 3 MeV electrons and annealed at 1050 $^\circ$C for two hours. [NV] is the concentration of negatively charged centers and its estimate is accurate to within a factor of two. [N] is the concentration of paramagnetic substitutional nitrogen. HPHT -- high-pressure high-temperature. CVD -- chemical vapor deposition. }
\centering
    \begin{ruledtabular}
    \begin{tabular}{c c c c c}
      \ Sample & Synthesis & Radiation dose & [N] & [NV] \\
      \  No. &  & (cm$^{-2}$) & (ppm) & (ppm)\\
      \colrule
      S2 & HPHT & $1 \times 10^{19}$ & $40-60$ & $16$ \\
      S8 & HPHT & $4 \times 10^{17}$ & $10-30$ & $0.3$ \\
      S3 & CVD & $4 \times 10^{17}$ & $0.1-4$ & $0.01$ \\
    \end{tabular}
    \end{ruledtabular}
\end{table}

The diamond samples used in this work are listed in Table \ref{tab:values}. These samples represent a wide range of nitrogen (N) and NV concentrations. Measurements were performed using a confocal-microscopy apparatus described previously \cite{ACO2010APL, ACO2010PRB}. An aspheric lens with focal length f = 6 mm and numerical aperture NA = 0.55 was used to focus 532 nm laser light on a diamond sample mounted in a flow-through liquid-helium cryostat. Fluorescence was collected by the same lens, passed through a dichroic mirror and was detected with a Si avalanche photodetector (Thorlabs APD110A) in the range of 650-800 nm. A microwave (MW) field was applied to the NV ensemble by means of a copper wire, 75 $\mu$m in diameter, placed inside the cryostat in the vicinity of the optical focus. Light pulses were generated by passing cw laser light through two acousto-optical modulators, which provided a combined extinction ratio of about 60 dB. MW pulses were produced by two series-connected switches, providing a combined isolation of $\sim$ 120 dB. High isolation is critical, particularly for measurements of very long $T_{1}$, in order to avoid unintended changes in polarization from stray fields. Sequences of pulses were produced by a programmable pulse generator PulseBlasterESR-PRO.

\begin{figure}
\centering
    \includegraphics[width=1.0\columnwidth]{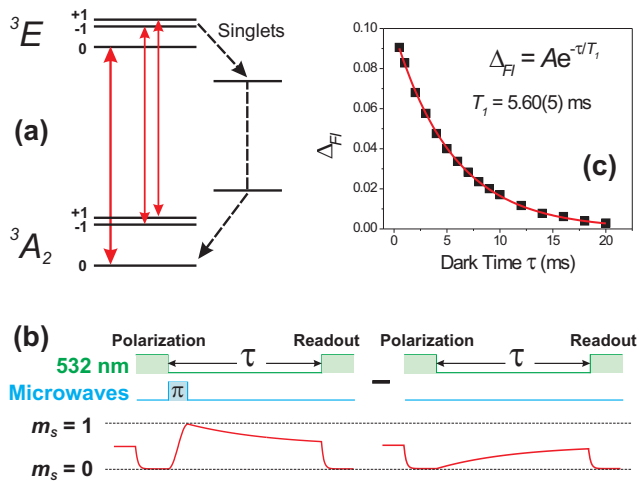}
    \caption{\label{fig:method} (a) Level structure of the NV center showing spin-triplet ground $^{3}A_{2}$ and excited $^{3}E$ states, as well as the singlet system involved in the optical spin-polarization mechanism. (b) Optical and microwave pulse sequence used for $T_{1}$ measurements. (c) Example of $T_{1}$ measurement. The difference in level of fluorescence for two measurements with and without a MW $\pi$-pulse, $\Delta_{Fl}$, is plotted as a function of dark time $\tau$ and fit to single-exponential decay.}
\end{figure}

Optical pumping and detection of spin polarization were performed using 532 nm excitation. Optical pumping occurs due to a spin-dependent intersystem crossing which transfers NV centers with spin projection, $m_s=\pm1$, to $m_s=0$ ground-state sublevel [Fig. \ref{fig:method}(a)]. Due to the same process, NV centers in $m_s=0$, fluoresce more brightly than NV centers in $m_s=\pm1$, allowing us to use the detected fluorescence intensity to determine the ground-state spin polarization.

Figure \ref{fig:method}(b) shows the timing sequence used in the experiments to measure $T_{1}$. Each sequence begins with a 1 ms light pulse which polarizes NV centers into $m_s=0$. We then apply a MW $\pi$-pulse, with typical duration on the order of 100 ns, which transfers NV centers into $m_s=1$ or $m_s=-1$. Following a variable time delay, $\tau$, we apply another light pulse and detect NV fluorescence to determine the residual spin polarization. Afterwards we apply a control sequence with the $\pi$-pulse omitted, and subtract the two results. This technique results in a signal, $\Delta_{Fl}(\tau)$, which is proportional to the residual spin polarization, after time $\tau$, of only those NV centers which were excited by the MW $\pi$-pulse. Figure 1(c) shows a typical plot of $\Delta_{Fl}$ versus $\tau$ at T = 300 K, along with a fit to single-exponential function, $\Delta_{Fl}(\tau)=A e^{-\tau/T_1}$. We found that all measurements reported in this manuscript can be satisfactorily fitted with a single decay constant.

\begin{figure}
\centering
    \includegraphics[width=1.0\columnwidth]{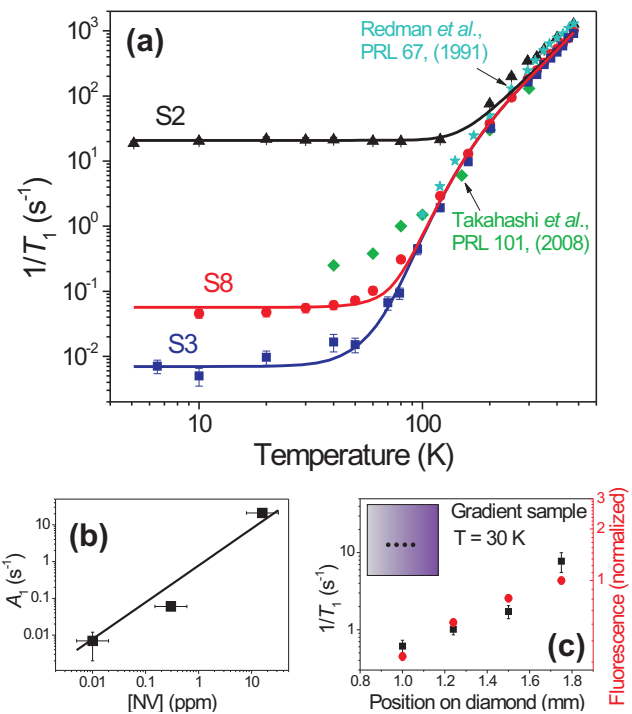}
    \caption{\label{fig:temp_dep} (a) $1/T_{1}$ as a function of temperature for S2, S3, and S8 samples. Symbols represent experimental data while solid lines are global fits to the model in Eq. (\ref{eq:T1}). Error bars represent the standard errors from single-exponential fits. For consistency, we also plot previously reported experimental data \cite{RED1991, TAK2008}. (b) $A_{1}$ [Eq. (\ref{eq:T1})] versus [NV]. (c) $1/T_{1}$ (squares) and fluorescence intensity (circles) at 30 K versus position on the sample with a spatially-varying electron irradiation dose. A magnetic field, B $\approx$ 25 G was applied along [100] direction for all measurements presented in this figure.}
\end{figure}

Insight into the nature of the longitudinal spin relaxation mechanisms can be gathered from its temperature dependence. Figure \ref{fig:temp_dep} shows the relaxation rate $1/T_{1}$ as a function of temperature for the three samples in Table \ref{tab:values}. A magnetic field, B $\approx$ 25 G was applied parallel to the [100] crystallographic orientation in order to separate $m_{s}=0 \rightarrow m_{s}=1$ and $m_{s}=0 \rightarrow m_{s}=-1$ transitions. At temperatures T > 200 K, $1/T_{1}$ is the same for all samples within a factor of $\approx$ 2. For example, at 300 K, we measure $T_{1}$ = 2.9(1), 5.5(1), and 6.0(1) ms for S2, S8, and S3, respectively. This suggests $T_{1}$ in this range is largely governed by intrinsic processes such as interaction with phonons \cite{SUP2012}. In contrast, at low temperatures, a strong sample dependence of $T_{1}$ is observed. The relaxation rate $1/T_{1}$ rapidly decreases with decreasing temperature and flattens out at a certain value which is different for each sample.

Considering interaction with lattice phonons, we note that the NV ground-state spin splitting, $D\approx$ 3 GHz. Consequently for T $\gg D/k \approx$ 0.1 K, where $k$ is the Boltzmann constant, the occupation of phonon modes near 3 GHz is extremely low, and we can neglect relaxation from a single-phonon process. However, higher-energy lattice phonons can produce $T_{1}$ relaxation via a two-phonon Raman process. We expect this process to have the form $(1/T_{1})_\mathrm{Raman} \propto T^{5}$ \cite{WAL1968}. Similarly, local phonons can produce $T_{1}$ relaxation by resonant interaction with an excited vibrational level via a two-phonon Orbach-type process \cite{RED1991}. In this case, we expect $(1/T_{1})_\mathrm{Orbach} \propto (e^{\frac{\Delta}{kT}}-1)^{-1}$, where $\Delta$ is the dominant local vibrational energy.

Based on Fig. \ref{fig:temp_dep}(a), we also observe a leveling-off of $(1/T_{1})$ at low temperatures to a constant limiting rate. This rate is sample dependent and represents, as it will be discussed later, cross relaxation which arises from dipole-dipole interaction. We fit our data to the following global function:

\begin{equation}
\label{eq:T1}
 \frac{1}{T_{1}}=A_{1}(\mathrm{S})+\frac{A_{2}}{e^{\frac{\Delta}{kT}}-1}+A_{3}T^{5},
\end{equation}
where $A_{2}$, $A_{3}$, and $\Delta$ are fit parameters common to all samples, and $A_{1}$(S) is different for each sample. The fit is in excellent agreement with experimental data (see \cite{SUP2012} for a comparison with the fit parameters individual for each sample). The values for the fitting parameters are: $A_{1}$(S2) = 21(3) s$^{-1}$, $A_{1}$(S8) = 0.06(1) s$^{-1}$, $A_{1}$(S3) = 0.007(4) s$^{-1}$, $A_{2}$ = 2.1(6) $\times$ 10$^{3}$ s$^{-1}$, $\Delta$ = 73(4) meV, and $A_{3}$ = 2.2(5) $\times$ 10$^{-11}$ K$^{-5}$s$^{-1}$. The value of the parameter $\Delta$ closely matches the energy of the local vibrational modes of the NV center \cite{ZHA2011, GAL2011}.

Figure \ref{fig:temp_dep}(b) shows $A_{1}$ as a function of NV concentration. The observed dependence is approximately linear, and the fitted slope is 0.8(2) s$^{-1}$ppm$^{-1}$. To verify this observation we performed $T_{1}$ measurements on a HPHT sample which has a spatial gradient in electron-irradiation dose and, correspondingly, in NV concentration. The nitrogen concentration of this diamond was measured as in Ref. \cite{ACO2009} and found to be on the order of 100 ppm throughout the sample. We measured $T_{1}$ at T = 30 K in four different spots along the dose gradient. The relaxation rate $1/T_{1}$ and fluorescence intensity for each position on the diamond are presented in Fig. \ref{fig:temp_dep}(c). Making an assumption that fluorescence intensity is proportional to NV concentration the results are consistent with linear dependence of $A_{1}$ on [NV].

\begin{figure}
\centering
    \includegraphics[width=1.0\columnwidth]{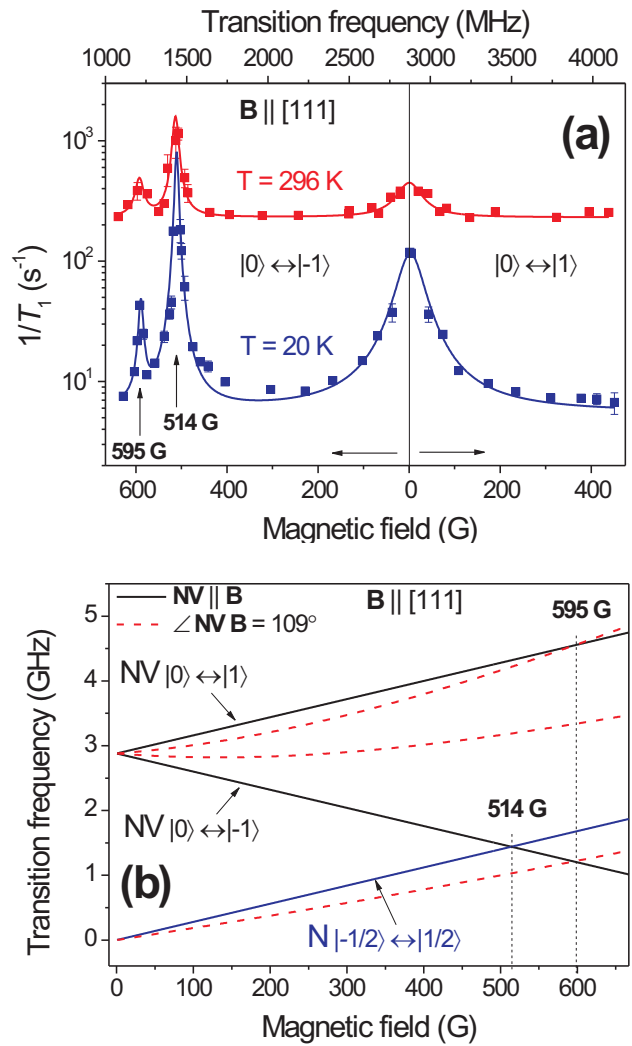}
    \caption{\label{fig:mag_dep} (a) $1/T_{1}$ as a function of the magnetic field $\textbf{B}$ applied along [111] direction (the corresponding transition frequencies are shown on the upper axis) for S2 sample at two different temperatures 296 K and 20 K. $1/T_{1}$ was measured for the subensemble of NV centers aligned with $\textbf{B}$. Solid line represents a fit (see text). (b) Transition frequencies between magnetic sublevels of NV centers and between sublevels of N as a function of magnetic field along [111] direction. NV-NV cross relaxation occurs at 0 G and 595 G, NV-N - at 514 G. Solid lines correspond to the centers aligned with [111]; dashed lines correspond to the NV centers aligned at an $\approx$ 109$^{\circ}$ to this axis.}
\end{figure}

In order to gain further insight into the apparent cross relaxation processes, we measured magnetic field dependence of $T_{1}$ for the sample with the highest impurity concentration, S2. Magnetic field $\textbf{B}$ was applied along the [111] axis using a neodymium ring magnet mounted on a 3-axis translation stage. For this magnetic field direction, one of four possible NV orientations is aligned parallel to $\textbf{B}$, while the other three orientations have symmetry axes which are at $\approx$ 109$^{\circ}$ with respect to $\textbf{B}$.  An additional 3-axis magnetic-coil system was used for fine tuning and scanning of the magnetic field.

Figure \ref{fig:mag_dep} (a) shows $1/T_{1}$ as a function of magnetic field for the subensemble of NV centers aligned along the magnetic field. The upper axis shows the corresponding MW transition frequencies that were probed. The data were taken at two different temperatures: 20 K and 296 K. Significant changes in $1/T_{1}$ are observed at certain magnetic field values: 0, 514, and 595 G. The relative changes are more pronounced for low-temperature data, where $1/T_{1}$ varies by more than an order of magnitude.

Cross relaxation occurs when the transition frequency of two spin species, each having different spin polarizations, coincide. Transition frequencies as a function of $\textbf{B}$ of the two sub-ensembles of NV centers (those co-linear and non-colinear with $\textbf{B}$) as well as that of the spin-1/2 nitrogen impurities are shown in Fig. \ref{fig:mag_dep} (b). The 514 G peak apparently arises due to cross relaxation between NV centers co-aligned with $\textbf{B}$ and the unpolarized nitrogen spins. At 0 and 595 G, the transition frequencies between the two subensembles of NV centers coincide. Cross relaxation can then occur if the subensembles are polarized differently by the initial optical pulse.  For example, the peak at 595 G (1.2 GHz) occurs due to cross relaxation between $|m_s=0\rangle \rightarrow |m_s=-1\rangle$ transition of NV centers aligned along the field direction and transitions between the eigenstates of other three degenerate orientations. At this field, each of the eigenstates is a mixture of the initial magnetic sublevels due to the large transverse field component. Cross relaxation at 0 G may be due to spin diffusion involving NV centers outside of the optical focus. Similar cross relaxation phenomena were previously studied for NV centers in diamond using other techniques, for example, monitoring changes of emission intensity \cite{VAN1989, VAN1991AMR, ARM2010}, measuring changes of the optical hole depth in the zero-phonon line \cite{HOL1989}, and studying decay rate of stimulated spin-echo and spin-locking signals \cite{VAN1989}.

We fit our experimental results to three Lorentzians with central frequencies fixed at 0, 514, and 595 G \footnote{Hyperfine structure associated with $^{14}$N is not resolved in the present experiments.}. The amplitude at 20 K is a factor of two smaller than at 296 K for 0 and 514 G peaks, and a factor of five smaller for the 595 G peak. We also observe a factor of two decrease in width for the 514 and 595 G peaks at 20 K. We did not expect to see a temperature dependence of either amplitudes or widths (since dipolar coupling strengths are not expected to be temperature dependent in this range \cite{TAK2008}); so the observed difference is a subject for future work. At both temperatures, the peak at 514 G is the largest peak, which is consistent with the [N]:[NV] ratio in Table \ref{tab:values}.

Let us look at the results from the point of view of magnetic sensing. Assuming that, ultimately, transverse relaxation times $T_{2}$ on the order of $T_{1}$ can be achieved \cite{DEL2010SCIENCE, RYA2010, NAY2011, PHA2012}, (which may require the use of diamond samples depleted of $^{13}$C \cite{MIZ2009} and dynamic decoupling
from the spin bath), the product [NV]$T_{1}$  is the figure of merit in terms of the sensor sensitivity \cite{TAY2008}. Examining the data in Fig. \ref{fig:temp_dep}(a), we find that at the lowest temperatures, [NV]$T_{1}$ is roughly independent of concentration, owing to the aforementioned NV-NV cross relaxation. At higher temperatures, the relaxation rate is dominated by intrinsic phonon interactions and consequently independent of [NV]. Thus the product [NV]$T_{1}$  increases linearly with concentration at a given temperature in this range. This motivates the use of high-density samples in room-temperature applications. It is important to notice that, with appropriate choice of magnetic field, it is possible to "tune" the NV centers to a point with an optimum value of $T_{1}$. For example, applying bias field to a low-temperature high-density sample, it is possible to prolong $T_{1}$ by more than an order of magnitude compared to near-zero field (Fig. \ref{fig:mag_dep} (a)). This property may also benefit other applications, including spin-ensemble memories in hybrid devices \cite{KUB2011, KUB2010, AMS2011, ZHU2011}.

In conclusion, we have investigated the temperature and magnetic-field dependence of longitudinal spin relaxation of NV ensembles in diamond. We observed two distinct regimes (low and high temperature), where different sets of relaxation mechanisms dominate. The dominant relaxation mechanism at high temperature is consistent with the exponential behavior of the two-phonon Orbach-type process and T$^5$ behavior of the two-phonon Raman process. At low temperatures $T_{1}$ is temperature independent and the relaxation is mainly determined by cross relaxation with neighboring spins. Through our magnetic-field studies, we identify this cross relaxation as being between differently polarized NV centers as well as between NV centers and nitrogen impurities. The longest $T_{1}$ observed was 199(41) s at 10 K for the CVD sample with low NV concentration. Our studies inform on the fundamental quantum limits of NV based computing and sensing applications.

The authors thank M. Ledbetter, N. Yao, A. Gali, N. Manson, P. Hemmer, L-S. Bouchard, W. Gawlik, M.W. Doherty, R. Fischer, P. G. Baranov, and T. Schenkel for useful discussions. This work was supported by NSF Grant No. PHY-0855552, AFOSR/DARPA QuASAR program, IMOD, and the NATO SFP program. A.J. gratefully acknowledges support from the ERAF project No. 2010/0242/2DP/2.1.1.1.0/10/APIA/VIAA/036.


\begin{thebibliography}{39}%
\makeatletter
\providecommand \@ifxundefined [1]{%
 \@ifx{#1\undefined}
}%
\providecommand \@ifnum [1]{%
 \ifnum #1\expandafter \@firstoftwo
 \else \expandafter \@secondoftwo
 \fi
}%
\providecommand \@ifx [1]{%
 \ifx #1\expandafter \@firstoftwo
 \else \expandafter \@secondoftwo
 \fi
}%
\providecommand \natexlab [1]{#1}%
\providecommand \enquote  [1]{``#1''}%
\providecommand \bibnamefont  [1]{#1}%
\providecommand \bibfnamefont [1]{#1}%
\providecommand \citenamefont [1]{#1}%
\providecommand \href@noop [0]{\@secondoftwo}%
\providecommand \href [0]{\begingroup \@sanitize@url \@href}%
\providecommand \@href[1]{\@@startlink{#1}\@@href}%
\providecommand \@@href[1]{\endgroup#1\@@endlink}%
\providecommand \@sanitize@url [0]{\catcode `\\12\catcode `\$12\catcode
  `\&12\catcode `\#12\catcode `\^12\catcode `\_12\catcode `\%12\relax}%
\providecommand \@@startlink[1]{}%
\providecommand \@@endlink[0]{}%
\providecommand \url  [0]{\begingroup\@sanitize@url \@url }%
\providecommand \@url [1]{\endgroup\@href {#1}{\urlprefix }}%
\providecommand \urlprefix  [0]{URL }%
\providecommand \Eprint [0]{\href }%
\providecommand \doibase [0]{http://dx.doi.org/}%
\providecommand \selectlanguage [0]{\@gobble}%
\providecommand \bibinfo  [0]{\@secondoftwo}%
\providecommand \bibfield  [0]{\@secondoftwo}%
\providecommand \translation [1]{[#1]}%
\providecommand \BibitemOpen [0]{}%
\providecommand \bibitemStop [0]{}%
\providecommand \bibitemNoStop [0]{.\EOS\space}%
\providecommand \EOS [0]{\spacefactor3000\relax}%
\providecommand \BibitemShut  [1]{\csname bibitem#1\endcsname}%
\let\auto@bib@innerbib\@empty
\bibitem [{\citenamefont {Fu}\ \emph {et~al.}(2007)\citenamefont {Fu},
  \citenamefont {Lee}, \citenamefont {Chen}, \citenamefont {Lim}, \citenamefont
  {Wu}, \citenamefont {Lin}, \citenamefont {Wei}, \citenamefont {Tsao},
  \citenamefont {Chang},\ and\ \citenamefont {Fann}}]{FU2007}%
  \BibitemOpen
  \bibfield  {author} {\bibinfo {author} {\bibfnamefont {C.-C.}\ \bibnamefont
  {Fu}}, \bibinfo {author} {\bibfnamefont {H.-Y.}\ \bibnamefont {Lee}},
  \bibinfo {author} {\bibfnamefont {K.}~\bibnamefont {Chen}}, \bibinfo {author}
  {\bibfnamefont {T.-S.}\ \bibnamefont {Lim}}, \bibinfo {author} {\bibfnamefont
  {H.-Y.}\ \bibnamefont {Wu}}, \bibinfo {author} {\bibfnamefont {P.-K.}\
  \bibnamefont {Lin}}, \bibinfo {author} {\bibfnamefont {P.-K.}\ \bibnamefont
  {Wei}}, \bibinfo {author} {\bibfnamefont {P.-H.}\ \bibnamefont {Tsao}},
  \bibinfo {author} {\bibfnamefont {H.-C.}\ \bibnamefont {Chang}}, \ and\
  \bibinfo {author} {\bibfnamefont {W.}~\bibnamefont {Fann}},\ }\href@noop {}
  {\bibfield  {journal} {\bibinfo  {journal} {Proceedings of the National
  Academy of Sciences}\ }\textbf {\bibinfo {volume} {104}},\ \bibinfo {pages}
  {727} (\bibinfo {year} {2007})}\BibitemShut {NoStop}%
\bibitem [{\citenamefont {Faklaris}\ \emph {et~al.}(2009)\citenamefont
  {Faklaris}, \citenamefont {Joshi}, \citenamefont {Irinopoulou}, \citenamefont
  {Tauc}, \citenamefont {Sennour}, \citenamefont {Girard}, \citenamefont
  {Gesset}, \citenamefont {Arnault}, \citenamefont {Thorel}, \citenamefont
  {Boudou}, \citenamefont {Curmi},\ and\ \citenamefont {Treussart}}]{FAK2009}%
  \BibitemOpen
  \bibfield  {author} {\bibinfo {author} {\bibfnamefont {O.}~\bibnamefont
  {Faklaris}}, \bibinfo {author} {\bibfnamefont {V.}~\bibnamefont {Joshi}},
  \bibinfo {author} {\bibfnamefont {T.}~\bibnamefont {Irinopoulou}}, \bibinfo
  {author} {\bibfnamefont {P.}~\bibnamefont {Tauc}}, \bibinfo {author}
  {\bibfnamefont {M.}~\bibnamefont {Sennour}}, \bibinfo {author} {\bibfnamefont
  {H.}~\bibnamefont {Girard}}, \bibinfo {author} {\bibfnamefont
  {C.}~\bibnamefont {Gesset}}, \bibinfo {author} {\bibfnamefont {J.-C.}\
  \bibnamefont {Arnault}}, \bibinfo {author} {\bibfnamefont {A.}~\bibnamefont
  {Thorel}}, \bibinfo {author} {\bibfnamefont {J.-P.}\ \bibnamefont {Boudou}},
  \bibinfo {author} {\bibfnamefont {P.~A.}\ \bibnamefont {Curmi}}, \ and\
  \bibinfo {author} {\bibfnamefont {F.}~\bibnamefont {Treussart}},\ }\href
  {\doibase 10.1021/nn901014j} {\bibfield  {journal} {\bibinfo  {journal} {ACS
  Nano}\ }\textbf {\bibinfo {volume} {3}},\ \bibinfo {pages} {3955} (\bibinfo
  {year} {2009})}\BibitemShut {NoStop}%
\bibitem [{\citenamefont {McGuinness}\ \emph {et~al.}(2011)\citenamefont
  {McGuinness}, \citenamefont {Yan}, \citenamefont {Stacey}, \citenamefont
  {Simpson}, \citenamefont {Hall}, \citenamefont {Maclaurin}, \citenamefont
  {Prawer}, \citenamefont {Mulvaney}, \citenamefont {Wrachtrup}, \citenamefont
  {Caruso}, \citenamefont {Scholten},\ and\ \citenamefont
  {Hollenberg}}]{MCG2011}%
  \BibitemOpen
  \bibfield  {author} {\bibinfo {author} {\bibfnamefont {L.~P.}\ \bibnamefont
  {McGuinness}}, \bibinfo {author} {\bibfnamefont {Y.}~\bibnamefont {Yan}},
  \bibinfo {author} {\bibfnamefont {A.}~\bibnamefont {Stacey}}, \bibinfo
  {author} {\bibfnamefont {D.~A.}\ \bibnamefont {Simpson}}, \bibinfo {author}
  {\bibfnamefont {L.~T.}\ \bibnamefont {Hall}}, \bibinfo {author}
  {\bibfnamefont {D.}~\bibnamefont {Maclaurin}}, \bibinfo {author}
  {\bibfnamefont {S.}~\bibnamefont {Prawer}}, \bibinfo {author} {\bibfnamefont
  {P.}~\bibnamefont {Mulvaney}}, \bibinfo {author} {\bibfnamefont
  {J.}~\bibnamefont {Wrachtrup}}, \bibinfo {author} {\bibfnamefont
  {F.}~\bibnamefont {Caruso}}, \bibinfo {author} {\bibfnamefont {R.~E.}\
  \bibnamefont {Scholten}}, \ and\ \bibinfo {author} {\bibfnamefont {L.~C.~L.}\
  \bibnamefont {Hollenberg}},\ }\href@noop {} {\bibfield  {journal} {\bibinfo
  {journal} {Nat Nano}\ }\textbf {\bibinfo {volume} {6}},\ \bibinfo {pages}
  {358} (\bibinfo {year} {2011})}\BibitemShut {NoStop}%
\bibitem [{\citenamefont {Neumann}\ \emph {et~al.}(2010)\citenamefont
  {Neumann}, \citenamefont {Kolesov}, \citenamefont {Naydenov}, \citenamefont
  {Beck}, \citenamefont {Rempp}, \citenamefont {Steiner}, \citenamefont
  {Jacques}, \citenamefont {Balasubramanian}, \citenamefont {Markham},
  \citenamefont {Twitchen}, \citenamefont {Pezzagna}, \citenamefont {Meijer},
  \citenamefont {Twamley}, \citenamefont {Jelezko},\ and\ \citenamefont
  {Wrachtrup}}]{NEU2010}%
  \BibitemOpen
  \bibfield  {author} {\bibinfo {author} {\bibfnamefont {P.}~\bibnamefont
  {Neumann}}, \bibinfo {author} {\bibfnamefont {R.}~\bibnamefont {Kolesov}},
  \bibinfo {author} {\bibfnamefont {B.}~\bibnamefont {Naydenov}}, \bibinfo
  {author} {\bibfnamefont {J.}~\bibnamefont {Beck}}, \bibinfo {author}
  {\bibfnamefont {F.}~\bibnamefont {Rempp}}, \bibinfo {author} {\bibfnamefont
  {M.}~\bibnamefont {Steiner}}, \bibinfo {author} {\bibfnamefont
  {V.}~\bibnamefont {Jacques}}, \bibinfo {author} {\bibfnamefont
  {G.}~\bibnamefont {Balasubramanian}}, \bibinfo {author} {\bibfnamefont
  {M.~L.}\ \bibnamefont {Markham}}, \bibinfo {author} {\bibfnamefont {D.~J.}\
  \bibnamefont {Twitchen}}, \bibinfo {author} {\bibfnamefont {S.}~\bibnamefont
  {Pezzagna}}, \bibinfo {author} {\bibfnamefont {J.}~\bibnamefont {Meijer}},
  \bibinfo {author} {\bibfnamefont {J.}~\bibnamefont {Twamley}}, \bibinfo
  {author} {\bibfnamefont {F.}~\bibnamefont {Jelezko}}, \ and\ \bibinfo
  {author} {\bibfnamefont {J.}~\bibnamefont {Wrachtrup}},\ }\href@noop {}
  {\bibfield  {journal} {\bibinfo  {journal} {Nature Physics}\ }\textbf
  {\bibinfo {volume} {6}},\ \bibinfo {pages} {249} (\bibinfo {year}
  {2010})}\BibitemShut {NoStop}%
\bibitem [{\citenamefont {Togan}\ \emph {et~al.}(2010)\citenamefont {Togan},
  \citenamefont {Chu}, \citenamefont {Trifonov}, \citenamefont {Jiang},
  \citenamefont {Maze}, \citenamefont {Childress}, \citenamefont {Dutt},
  \citenamefont {Sorensen}, \citenamefont {Hemmer}, \citenamefont {Zibrov},\
  and\ \citenamefont {Lukin}}]{TOG2010}%
  \BibitemOpen
  \bibfield  {author} {\bibinfo {author} {\bibfnamefont {E.}~\bibnamefont
  {Togan}}, \bibinfo {author} {\bibfnamefont {Y.}~\bibnamefont {Chu}}, \bibinfo
  {author} {\bibfnamefont {A.~S.}\ \bibnamefont {Trifonov}}, \bibinfo {author}
  {\bibfnamefont {L.}~\bibnamefont {Jiang}}, \bibinfo {author} {\bibfnamefont
  {J.}~\bibnamefont {Maze}}, \bibinfo {author} {\bibfnamefont {L.}~\bibnamefont
  {Childress}}, \bibinfo {author} {\bibfnamefont {M.~V.~G.}\ \bibnamefont
  {Dutt}}, \bibinfo {author} {\bibfnamefont {A.~S.}\ \bibnamefont {Sorensen}},
  \bibinfo {author} {\bibfnamefont {P.~R.}\ \bibnamefont {Hemmer}}, \bibinfo
  {author} {\bibfnamefont {A.~S.}\ \bibnamefont {Zibrov}}, \ and\ \bibinfo
  {author} {\bibfnamefont {M.~D.}\ \bibnamefont {Lukin}},\ }\href@noop {}
  {\bibfield  {journal} {\bibinfo  {journal} {Nature}\ }\textbf {\bibinfo
  {volume} {466}},\ \bibinfo {pages} {730} (\bibinfo {year}
  {2010})}\BibitemShut {NoStop}%
\bibitem [{\citenamefont {Robledo}\ \emph {et~al.}(2011)\citenamefont
  {Robledo}, \citenamefont {Childress}, \citenamefont {Bernien}, \citenamefont
  {Hensen}, \citenamefont {Alkemade},\ and\ \citenamefont
  {Hanson}}]{ROB2011NAT}%
  \BibitemOpen
  \bibfield  {author} {\bibinfo {author} {\bibfnamefont {L.}~\bibnamefont
  {Robledo}}, \bibinfo {author} {\bibfnamefont {L.}~\bibnamefont {Childress}},
  \bibinfo {author} {\bibfnamefont {H.}~\bibnamefont {Bernien}}, \bibinfo
  {author} {\bibfnamefont {B.}~\bibnamefont {Hensen}}, \bibinfo {author}
  {\bibfnamefont {P.~F.~A.}\ \bibnamefont {Alkemade}}, \ and\ \bibinfo {author}
  {\bibfnamefont {R.}~\bibnamefont {Hanson}},\ }\href@noop {} {\bibfield
  {journal} {\bibinfo  {journal} {Nature}\ }\textbf {\bibinfo {volume} {477}},\
  \bibinfo {pages} {574} (\bibinfo {year} {2011})}\BibitemShut {NoStop}%
\bibitem [{\citenamefont {Taylor}\ \emph {et~al.}(2008)\citenamefont {Taylor},
  \citenamefont {Cappellaro}, \citenamefont {Childress}, \citenamefont {Jiang},
  \citenamefont {Budker}, \citenamefont {Hemmer}, \citenamefont {Yacoby},
  \citenamefont {Walsworth},\ and\ \citenamefont {Lukin}}]{TAY2008}%
  \BibitemOpen
  \bibfield  {author} {\bibinfo {author} {\bibfnamefont {J.~M.}\ \bibnamefont
  {Taylor}}, \bibinfo {author} {\bibfnamefont {P.}~\bibnamefont {Cappellaro}},
  \bibinfo {author} {\bibfnamefont {L.}~\bibnamefont {Childress}}, \bibinfo
  {author} {\bibfnamefont {L.}~\bibnamefont {Jiang}}, \bibinfo {author}
  {\bibfnamefont {D.}~\bibnamefont {Budker}}, \bibinfo {author} {\bibfnamefont
  {P.~R.}\ \bibnamefont {Hemmer}}, \bibinfo {author} {\bibfnamefont
  {A.}~\bibnamefont {Yacoby}}, \bibinfo {author} {\bibfnamefont
  {R.}~\bibnamefont {Walsworth}}, \ and\ \bibinfo {author} {\bibfnamefont
  {M.~D.}\ \bibnamefont {Lukin}},\ }\href@noop {} {\bibfield  {journal}
  {\bibinfo  {journal} {Nat Phys}\ }\textbf {\bibinfo {volume} {4}},\ \bibinfo
  {pages} {810} (\bibinfo {year} {2008})}\BibitemShut {NoStop}%
\bibitem [{\citenamefont {Balasubramanian}\ \emph {et~al.}(2008)\citenamefont
  {Balasubramanian}, \citenamefont {Chan}, \citenamefont {Kolesov},
  \citenamefont {Al-Hmoud}, \citenamefont {Tisler}, \citenamefont {Shin},
  \citenamefont {Kim}, \citenamefont {Wojcik}, \citenamefont {Hemmer},
  \citenamefont {Krueger}, \citenamefont {Hanke}, \citenamefont
  {Leitenstorfer}, \citenamefont {Bratschitsch}, \citenamefont {Jelezko},\ and\
  \citenamefont {Wrachtrup}}]{BAL2008}%
  \BibitemOpen
  \bibfield  {author} {\bibinfo {author} {\bibfnamefont {G.}~\bibnamefont
  {Balasubramanian}}, \bibinfo {author} {\bibfnamefont {I.~Y.}\ \bibnamefont
  {Chan}}, \bibinfo {author} {\bibfnamefont {R.}~\bibnamefont {Kolesov}},
  \bibinfo {author} {\bibfnamefont {M.}~\bibnamefont {Al-Hmoud}}, \bibinfo
  {author} {\bibfnamefont {J.}~\bibnamefont {Tisler}}, \bibinfo {author}
  {\bibfnamefont {C.}~\bibnamefont {Shin}}, \bibinfo {author} {\bibfnamefont
  {C.}~\bibnamefont {Kim}}, \bibinfo {author} {\bibfnamefont {A.}~\bibnamefont
  {Wojcik}}, \bibinfo {author} {\bibfnamefont {P.~R.}\ \bibnamefont {Hemmer}},
  \bibinfo {author} {\bibfnamefont {A.}~\bibnamefont {Krueger}}, \bibinfo
  {author} {\bibfnamefont {T.}~\bibnamefont {Hanke}}, \bibinfo {author}
  {\bibfnamefont {A.}~\bibnamefont {Leitenstorfer}}, \bibinfo {author}
  {\bibfnamefont {R.}~\bibnamefont {Bratschitsch}}, \bibinfo {author}
  {\bibfnamefont {F.}~\bibnamefont {Jelezko}}, \ and\ \bibinfo {author}
  {\bibfnamefont {J.}~\bibnamefont {Wrachtrup}},\ }\href@noop {} {\bibfield
  {journal} {\bibinfo  {journal} {Nature}\ }\textbf {\bibinfo {volume} {455}},\
  \bibinfo {pages} {648} (\bibinfo {year} {2008})}\BibitemShut {NoStop}%
\bibitem [{\citenamefont {Maze}\ \emph {et~al.}(2008)\citenamefont {Maze},
  \citenamefont {Stanwix}, \citenamefont {Hodges}, \citenamefont {Hong},
  \citenamefont {Taylor}, \citenamefont {Cappellaro}, \citenamefont {Jiang},
  \citenamefont {Dutt}, \citenamefont {Togan}, \citenamefont {Zibrov},
  \citenamefont {Yacoby}, \citenamefont {Walsworth},\ and\ \citenamefont
  {Lukin}}]{MAZ2008NATURE}%
  \BibitemOpen
  \bibfield  {author} {\bibinfo {author} {\bibfnamefont {J.~R.}\ \bibnamefont
  {Maze}}, \bibinfo {author} {\bibfnamefont {P.~L.}\ \bibnamefont {Stanwix}},
  \bibinfo {author} {\bibfnamefont {J.~S.}\ \bibnamefont {Hodges}}, \bibinfo
  {author} {\bibfnamefont {S.}~\bibnamefont {Hong}}, \bibinfo {author}
  {\bibfnamefont {J.~M.}\ \bibnamefont {Taylor}}, \bibinfo {author}
  {\bibfnamefont {P.}~\bibnamefont {Cappellaro}}, \bibinfo {author}
  {\bibfnamefont {L.}~\bibnamefont {Jiang}}, \bibinfo {author} {\bibfnamefont
  {M.~V.~G.}\ \bibnamefont {Dutt}}, \bibinfo {author} {\bibfnamefont
  {E.}~\bibnamefont {Togan}}, \bibinfo {author} {\bibfnamefont {A.~S.}\
  \bibnamefont {Zibrov}}, \bibinfo {author} {\bibfnamefont {A.}~\bibnamefont
  {Yacoby}}, \bibinfo {author} {\bibfnamefont {R.~L.}\ \bibnamefont
  {Walsworth}}, \ and\ \bibinfo {author} {\bibfnamefont {M.~D.}\ \bibnamefont
  {Lukin}},\ }\href@noop {} {\bibfield  {journal} {\bibinfo  {journal}
  {Nature}\ }\textbf {\bibinfo {volume} {455}},\ \bibinfo {pages} {644}
  (\bibinfo {year} {2008})}\BibitemShut {NoStop}%
\bibitem [{\citenamefont {Acosta}\ \emph
  {et~al.}(2010{\natexlab{a}})\citenamefont {Acosta}, \citenamefont {Bauch},
  \citenamefont {Jarmola}, \citenamefont {Zipp}, \citenamefont {Ledbetter},\
  and\ \citenamefont {Budker}}]{ACO2010APL}%
  \BibitemOpen
  \bibfield  {author} {\bibinfo {author} {\bibfnamefont {V.~M.}\ \bibnamefont
  {Acosta}}, \bibinfo {author} {\bibfnamefont {E.}~\bibnamefont {Bauch}},
  \bibinfo {author} {\bibfnamefont {A.}~\bibnamefont {Jarmola}}, \bibinfo
  {author} {\bibfnamefont {L.~J.}\ \bibnamefont {Zipp}}, \bibinfo {author}
  {\bibfnamefont {M.~P.}\ \bibnamefont {Ledbetter}}, \ and\ \bibinfo {author}
  {\bibfnamefont {D.}~\bibnamefont {Budker}},\ }\href@noop {} {\bibfield
  {journal} {\bibinfo  {journal} {Applied Physics Letters}\ }\textbf {\bibinfo
  {volume} {97}},\ \bibinfo {pages} {174104} (\bibinfo {year}
  {2010}{\natexlab{a}})}\BibitemShut {NoStop}%
\bibitem [{\citenamefont {Dolde}\ \emph {et~al.}(2011)\citenamefont {Dolde},
  \citenamefont {Fedder}, \citenamefont {Doherty}, \citenamefont {Nobauer},
  \citenamefont {Rempp}, \citenamefont {Balasubramanian}, \citenamefont {Wolf},
  \citenamefont {Reinhard}, \citenamefont {Hollenberg}, \citenamefont
  {Jelezko},\ and\ \citenamefont {Wrachtrup}}]{DOL2011}%
  \BibitemOpen
  \bibfield  {author} {\bibinfo {author} {\bibfnamefont {F.}~\bibnamefont
  {Dolde}}, \bibinfo {author} {\bibfnamefont {H.}~\bibnamefont {Fedder}},
  \bibinfo {author} {\bibfnamefont {M.~W.}\ \bibnamefont {Doherty}}, \bibinfo
  {author} {\bibfnamefont {T.}~\bibnamefont {Nobauer}}, \bibinfo {author}
  {\bibfnamefont {F.}~\bibnamefont {Rempp}}, \bibinfo {author} {\bibfnamefont
  {G.}~\bibnamefont {Balasubramanian}}, \bibinfo {author} {\bibfnamefont
  {T.}~\bibnamefont {Wolf}}, \bibinfo {author} {\bibfnamefont {F.}~\bibnamefont
  {Reinhard}}, \bibinfo {author} {\bibfnamefont {L.~C.~L.}\ \bibnamefont
  {Hollenberg}}, \bibinfo {author} {\bibfnamefont {F.}~\bibnamefont {Jelezko}},
  \ and\ \bibinfo {author} {\bibfnamefont {J.}~\bibnamefont {Wrachtrup}},\
  }\href@noop {} {\bibfield  {journal} {\bibinfo  {journal} {Nature Physics}\
  }\textbf {\bibinfo {volume} {7}},\ \bibinfo {pages} {459} (\bibinfo {year}
  {2011})}\BibitemShut {NoStop}%
\bibitem [{\citenamefont {Acosta}\ \emph {et~al.}(2009)\citenamefont {Acosta},
  \citenamefont {Bauch}, \citenamefont {Ledbetter}, \citenamefont {Santori},
  \citenamefont {Fu}, \citenamefont {Barclay}, \citenamefont {Beausoleil},
  \citenamefont {Linget}, \citenamefont {Roch}, \citenamefont {Treussart},
  \citenamefont {Chemerisov}, \citenamefont {Gawlik},\ and\ \citenamefont
  {Budker}}]{ACO2009}%
  \BibitemOpen
  \bibfield  {author} {\bibinfo {author} {\bibfnamefont {V.~M.}\ \bibnamefont
  {Acosta}}, \bibinfo {author} {\bibfnamefont {E.}~\bibnamefont {Bauch}},
  \bibinfo {author} {\bibfnamefont {M.~P.}\ \bibnamefont {Ledbetter}}, \bibinfo
  {author} {\bibfnamefont {C.}~\bibnamefont {Santori}}, \bibinfo {author}
  {\bibfnamefont {K.~M.~C.}\ \bibnamefont {Fu}}, \bibinfo {author}
  {\bibfnamefont {P.~E.}\ \bibnamefont {Barclay}}, \bibinfo {author}
  {\bibfnamefont {R.~G.}\ \bibnamefont {Beausoleil}}, \bibinfo {author}
  {\bibfnamefont {H.}~\bibnamefont {Linget}}, \bibinfo {author} {\bibfnamefont
  {J.~F.}\ \bibnamefont {Roch}}, \bibinfo {author} {\bibfnamefont
  {F.}~\bibnamefont {Treussart}}, \bibinfo {author} {\bibfnamefont
  {S.}~\bibnamefont {Chemerisov}}, \bibinfo {author} {\bibfnamefont
  {W.}~\bibnamefont {Gawlik}}, \ and\ \bibinfo {author} {\bibfnamefont
  {D.}~\bibnamefont {Budker}},\ }\href@noop {} {\bibfield  {journal} {\bibinfo
  {journal} {Physical Review B}\ }\textbf {\bibinfo {volume} {80}},\ \bibinfo
  {pages} {115202} (\bibinfo {year} {2009})}\BibitemShut {NoStop}%
\bibitem [{\citenamefont {Maertz}\ \emph {et~al.}(2010)\citenamefont {Maertz},
  \citenamefont {Wijnheijmer}, \citenamefont {Fuchs}, \citenamefont
  {Nowakowski},\ and\ \citenamefont {Awschalom}}]{MAE2010}%
  \BibitemOpen
  \bibfield  {author} {\bibinfo {author} {\bibfnamefont {B.~J.}\ \bibnamefont
  {Maertz}}, \bibinfo {author} {\bibfnamefont {A.~P.}\ \bibnamefont
  {Wijnheijmer}}, \bibinfo {author} {\bibfnamefont {G.~D.}\ \bibnamefont
  {Fuchs}}, \bibinfo {author} {\bibfnamefont {M.~E.}\ \bibnamefont
  {Nowakowski}}, \ and\ \bibinfo {author} {\bibfnamefont {D.~D.}\ \bibnamefont
  {Awschalom}},\ }\href@noop {} {\bibfield  {journal} {\bibinfo  {journal}
  {Applied Physics Letters}\ }\textbf {\bibinfo {volume} {96}},\ \bibinfo
  {pages} {092504} (\bibinfo {year} {2010})}\BibitemShut {NoStop}%
\bibitem [{\citenamefont {Pham}\ \emph {et~al.}(2011)\citenamefont {Pham},
  \citenamefont {Le~Sage}, \citenamefont {Stanwix}, \citenamefont {Yeung},
  \citenamefont {Glenn}, \citenamefont {Trifonov}, \citenamefont {Cappellaro},
  \citenamefont {Hemmer}, \citenamefont {Lukin}, \citenamefont {Park},
  \citenamefont {Yacoby},\ and\ \citenamefont {Walsworth}}]{PHA2011}%
  \BibitemOpen
  \bibfield  {author} {\bibinfo {author} {\bibfnamefont {L.~M.}\ \bibnamefont
  {Pham}}, \bibinfo {author} {\bibfnamefont {D.}~\bibnamefont {Le~Sage}},
  \bibinfo {author} {\bibfnamefont {P.~L.}\ \bibnamefont {Stanwix}}, \bibinfo
  {author} {\bibfnamefont {T.~K.}\ \bibnamefont {Yeung}}, \bibinfo {author}
  {\bibfnamefont {D.}~\bibnamefont {Glenn}}, \bibinfo {author} {\bibfnamefont
  {A.}~\bibnamefont {Trifonov}}, \bibinfo {author} {\bibfnamefont
  {P.}~\bibnamefont {Cappellaro}}, \bibinfo {author} {\bibfnamefont {P.~R.}\
  \bibnamefont {Hemmer}}, \bibinfo {author} {\bibfnamefont {M.~D.}\
  \bibnamefont {Lukin}}, \bibinfo {author} {\bibfnamefont {H.}~\bibnamefont
  {Park}}, \bibinfo {author} {\bibfnamefont {A.}~\bibnamefont {Yacoby}}, \ and\
  \bibinfo {author} {\bibfnamefont {R.~L.}\ \bibnamefont {Walsworth}},\
  }\href@noop {} {\bibfield  {journal} {\bibinfo  {journal} {New Journal of
  Physics}\ }\textbf {\bibinfo {volume} {13}},\ \bibinfo {pages} {045021}
  (\bibinfo {year} {2011})}\BibitemShut {NoStop}%
\bibitem [{\citenamefont {DiVincenzo}(2000)}]{DIV2000}%
  \BibitemOpen
  \bibfield  {author} {\bibinfo {author} {\bibfnamefont {D.~P.}\ \bibnamefont
  {DiVincenzo}},\ }\href {\doibase
  10.1002/1521-3978(200009)48:9/11<771::AID-PROP771>3.0.CO;2-E} {\bibfield
  {journal} {\bibinfo  {journal} {Fortschritte der Physik}\ }\textbf {\bibinfo
  {volume} {48}},\ \bibinfo {pages} {771} (\bibinfo {year} {2000})}\BibitemShut
  {NoStop}%
\bibitem [{\citenamefont {Ladd}\ \emph {et~al.}(2010)\citenamefont {Ladd},
  \citenamefont {Jelezko}, \citenamefont {Laflamme}, \citenamefont {Nakamura},
  \citenamefont {Monroe},\ and\ \citenamefont {O'Brien}}]{LAD2010}%
  \BibitemOpen
  \bibfield  {author} {\bibinfo {author} {\bibfnamefont {T.~D.}\ \bibnamefont
  {Ladd}}, \bibinfo {author} {\bibfnamefont {F.}~\bibnamefont {Jelezko}},
  \bibinfo {author} {\bibfnamefont {R.}~\bibnamefont {Laflamme}}, \bibinfo
  {author} {\bibfnamefont {Y.}~\bibnamefont {Nakamura}}, \bibinfo {author}
  {\bibfnamefont {C.}~\bibnamefont {Monroe}}, \ and\ \bibinfo {author}
  {\bibfnamefont {J.~L.}\ \bibnamefont {O'Brien}},\ }\href@noop {} {\bibfield
  {journal} {\bibinfo  {journal} {Nature}\ }\textbf {\bibinfo {volume} {464}},\
  \bibinfo {pages} {45} (\bibinfo {year} {2010})}\BibitemShut {NoStop}%
\bibitem [{\citenamefont {Abragam}\ and\ \citenamefont
  {Bleaney}(1970)}]{ABR1970}%
  \BibitemOpen
  \bibfield  {author} {\bibinfo {author} {\bibfnamefont {A.}~\bibnamefont
  {Abragam}}\ and\ \bibinfo {author} {\bibfnamefont {B.}~\bibnamefont
  {Bleaney}},\ }\href@noop {} {\emph {\bibinfo {title} {Electron Paramagnetic
  Resonance of Transition Ions}}}\ (\bibinfo  {publisher} {Clarendon Press},\
  \bibinfo {address} {Oxford},\ \bibinfo {year} {1970})\BibitemShut {NoStop}%
\bibitem [{\citenamefont {de~Lange}\ \emph {et~al.}(2010)\citenamefont
  {de~Lange}, \citenamefont {Wang}, \citenamefont {Riste}, \citenamefont
  {Dobrovitski},\ and\ \citenamefont {Hanson}}]{DEL2010SCIENCE}%
  \BibitemOpen
  \bibfield  {author} {\bibinfo {author} {\bibfnamefont {G.}~\bibnamefont
  {de~Lange}}, \bibinfo {author} {\bibfnamefont {Z.~H.}\ \bibnamefont {Wang}},
  \bibinfo {author} {\bibfnamefont {D.}~\bibnamefont {Riste}}, \bibinfo
  {author} {\bibfnamefont {V.~V.}\ \bibnamefont {Dobrovitski}}, \ and\ \bibinfo
  {author} {\bibfnamefont {R.}~\bibnamefont {Hanson}},\ }\href@noop {}
  {\bibfield  {journal} {\bibinfo  {journal} {Science}\ }\textbf {\bibinfo
  {volume} {330}},\ \bibinfo {pages} {60} (\bibinfo {year} {2010})}\BibitemShut
  {NoStop}%
\bibitem [{\citenamefont {Ryan}\ \emph {et~al.}(2010)\citenamefont {Ryan},
  \citenamefont {Hodges},\ and\ \citenamefont {Cory}}]{RYA2010}%
  \BibitemOpen
  \bibfield  {author} {\bibinfo {author} {\bibfnamefont {C.~A.}\ \bibnamefont
  {Ryan}}, \bibinfo {author} {\bibfnamefont {J.~S.}\ \bibnamefont {Hodges}}, \
  and\ \bibinfo {author} {\bibfnamefont {D.~G.}\ \bibnamefont {Cory}},\
  }\href@noop {} {\bibfield  {journal} {\bibinfo  {journal} {Physical Review
  Letters}\ }\textbf {\bibinfo {volume} {105}},\ \bibinfo {pages} {200402}
  (\bibinfo {year} {2010})}\BibitemShut {NoStop}%
\bibitem [{\citenamefont {Naydenov}\ \emph {et~al.}(2011)\citenamefont
  {Naydenov}, \citenamefont {Dolde}, \citenamefont {Hall}, \citenamefont
  {Shin}, \citenamefont {Fedder}, \citenamefont {Hollenberg}, \citenamefont
  {Jelezko},\ and\ \citenamefont {Wrachtrup}}]{NAY2011}%
  \BibitemOpen
  \bibfield  {author} {\bibinfo {author} {\bibfnamefont {B.}~\bibnamefont
  {Naydenov}}, \bibinfo {author} {\bibfnamefont {F.}~\bibnamefont {Dolde}},
  \bibinfo {author} {\bibfnamefont {L.~T.}\ \bibnamefont {Hall}}, \bibinfo
  {author} {\bibfnamefont {C.}~\bibnamefont {Shin}}, \bibinfo {author}
  {\bibfnamefont {H.}~\bibnamefont {Fedder}}, \bibinfo {author} {\bibfnamefont
  {L.~C.~L.}\ \bibnamefont {Hollenberg}}, \bibinfo {author} {\bibfnamefont
  {F.}~\bibnamefont {Jelezko}}, \ and\ \bibinfo {author} {\bibfnamefont
  {J.}~\bibnamefont {Wrachtrup}},\ }\href@noop {} {\bibfield  {journal}
  {\bibinfo  {journal} {Physical Review B}\ }\textbf {\bibinfo {volume} {83}},\
  \bibinfo {pages} {081201} (\bibinfo {year} {2011})}\BibitemShut {NoStop}%
\bibitem [{\citenamefont {Pham}\ \emph {et~al.}(2012)\citenamefont {Pham},
  \citenamefont {Bar-Gill}, \citenamefont {Belthangady}, \citenamefont
  {Le~Sage}, \citenamefont {Cappellaro}, \citenamefont {Lukin}, \citenamefont
  {Yacoby},\ and\ \citenamefont {Walsworth}}]{PHA2012}%
  \BibitemOpen
  \bibfield  {author} {\bibinfo {author} {\bibfnamefont {L.~M.}\ \bibnamefont
  {Pham}}, \bibinfo {author} {\bibfnamefont {N.}~\bibnamefont {Bar-Gill}},
  \bibinfo {author} {\bibfnamefont {C.}~\bibnamefont {Belthangady}}, \bibinfo
  {author} {\bibfnamefont {D.}~\bibnamefont {Le~Sage}}, \bibinfo {author}
  {\bibfnamefont {P.}~\bibnamefont {Cappellaro}}, \bibinfo {author}
  {\bibfnamefont {M.~D.}\ \bibnamefont {Lukin}}, \bibinfo {author}
  {\bibfnamefont {A.}~\bibnamefont {Yacoby}}, \ and\ \bibinfo {author}
  {\bibfnamefont {R.~L.}\ \bibnamefont {Walsworth}},\ }\href@noop {} {\enquote
  {\bibinfo {title} {Enhanced solid-state multi-spin metrology using dynamical
  decoupling},}\ } (\bibinfo {year} {2012}),\ \bibinfo {note}
  {arXiv:1201.5686}\BibitemShut {NoStop}%
\bibitem [{\citenamefont {Redman}\ \emph {et~al.}(1991)\citenamefont {Redman},
  \citenamefont {Brown}, \citenamefont {Sands},\ and\ \citenamefont
  {Rand}}]{RED1991}%
  \BibitemOpen
  \bibfield  {author} {\bibinfo {author} {\bibfnamefont {D.~A.}\ \bibnamefont
  {Redman}}, \bibinfo {author} {\bibfnamefont {S.}~\bibnamefont {Brown}},
  \bibinfo {author} {\bibfnamefont {R.~H.}\ \bibnamefont {Sands}}, \ and\
  \bibinfo {author} {\bibfnamefont {S.~C.}\ \bibnamefont {Rand}},\ }\href@noop
  {} {\bibfield  {journal} {\bibinfo  {journal} {Physical Review Letters}\
  }\textbf {\bibinfo {volume} {67}},\ \bibinfo {pages} {3420} (\bibinfo {year}
  {1991})}\BibitemShut {NoStop}%
\bibitem [{\citenamefont {Takahashi}\ \emph {et~al.}(2008)\citenamefont
  {Takahashi}, \citenamefont {Hanson}, \citenamefont {van Tol}, \citenamefont
  {Sherwin},\ and\ \citenamefont {Awschalom}}]{TAK2008}%
  \BibitemOpen
  \bibfield  {author} {\bibinfo {author} {\bibfnamefont {S.}~\bibnamefont
  {Takahashi}}, \bibinfo {author} {\bibfnamefont {R.}~\bibnamefont {Hanson}},
  \bibinfo {author} {\bibfnamefont {J.}~\bibnamefont {van Tol}}, \bibinfo
  {author} {\bibfnamefont {M.~S.}\ \bibnamefont {Sherwin}}, \ and\ \bibinfo
  {author} {\bibfnamefont {D.~D.}\ \bibnamefont {Awschalom}},\ }\href@noop {}
  {\bibfield  {journal} {\bibinfo  {journal} {Physical Review Letters}\
  }\textbf {\bibinfo {volume} {101}},\ \bibinfo {pages} {047601} (\bibinfo
  {year} {2008})}\BibitemShut {NoStop}%
\bibitem [{\citenamefont {Harrison}\ \emph {et~al.}(2006)\citenamefont
  {Harrison}, \citenamefont {Sellars},\ and\ \citenamefont {Manson}}]{HAR2006}%
  \BibitemOpen
  \bibfield  {author} {\bibinfo {author} {\bibfnamefont {J.}~\bibnamefont
  {Harrison}}, \bibinfo {author} {\bibfnamefont {M.~J.}\ \bibnamefont
  {Sellars}}, \ and\ \bibinfo {author} {\bibfnamefont {N.~B.}\ \bibnamefont
  {Manson}},\ }\href@noop {} {\bibfield  {journal} {\bibinfo  {journal}
  {Diamond and Related Materials}\ }\textbf {\bibinfo {volume} {15}},\ \bibinfo
  {pages} {586} (\bibinfo {year} {2006})}\BibitemShut {NoStop}%
\bibitem [{\citenamefont {Acosta}\ \emph
  {et~al.}(2010{\natexlab{b}})\citenamefont {Acosta}, \citenamefont {Jarmola},
  \citenamefont {Bauch},\ and\ \citenamefont {Budker}}]{ACO2010PRB}%
  \BibitemOpen
  \bibfield  {author} {\bibinfo {author} {\bibfnamefont {V.~M.}\ \bibnamefont
  {Acosta}}, \bibinfo {author} {\bibfnamefont {A.}~\bibnamefont {Jarmola}},
  \bibinfo {author} {\bibfnamefont {E.}~\bibnamefont {Bauch}}, \ and\ \bibinfo
  {author} {\bibfnamefont {D.}~\bibnamefont {Budker}},\ }\href@noop {}
  {\bibfield  {journal} {\bibinfo  {journal} {Physical Review B}\ }\textbf
  {\bibinfo {volume} {82}},\ \bibinfo {pages} {201202} (\bibinfo {year}
  {2010}{\natexlab{b}})}\BibitemShut {NoStop}%
\bibitem [{SUP()}]{SUP2012}%
  \BibitemOpen
  \href@noop {} {}\bibinfo {note} {See Supplemental Material for details
  regarding data analysys and interpretation.}\BibitemShut {Stop}%
\bibitem [{\citenamefont {Walker}(1968)}]{WAL1968}%
  \BibitemOpen
  \bibfield  {author} {\bibinfo {author} {\bibfnamefont {M.~B.}\ \bibnamefont
  {Walker}},\ }\href {\doibase 10.1139/p68-455} {\bibfield  {journal} {\bibinfo
   {journal} {Canadian Journal of Physics}\ }\textbf {\bibinfo {volume} {46}},\
  \bibinfo {pages} {1347} (\bibinfo {year} {1968})}\BibitemShut {NoStop}%
\bibitem [{\citenamefont {Zhang}\ \emph {et~al.}(2011)\citenamefont {Zhang},
  \citenamefont {Wang}, \citenamefont {Zhu},\ and\ \citenamefont
  {Dobrovitski}}]{ZHA2011}%
  \BibitemOpen
  \bibfield  {author} {\bibinfo {author} {\bibfnamefont {J.}~\bibnamefont
  {Zhang}}, \bibinfo {author} {\bibfnamefont {C.-Z.}\ \bibnamefont {Wang}},
  \bibinfo {author} {\bibfnamefont {Z.~Z.}\ \bibnamefont {Zhu}}, \ and\
  \bibinfo {author} {\bibfnamefont {V.~V.}\ \bibnamefont {Dobrovitski}},\
  }\href {\doibase 10.1103/PhysRevB.84.035211} {\bibfield  {journal} {\bibinfo
  {journal} {Phys. Rev. B}\ }\textbf {\bibinfo {volume} {84}},\ \bibinfo
  {pages} {035211} (\bibinfo {year} {2011})}\BibitemShut {NoStop}%
\bibitem [{\citenamefont {Gali}\ \emph {et~al.}(2011)\citenamefont {Gali},
  \citenamefont {Simon},\ and\ \citenamefont {Lowther}}]{GAL2011}%
  \BibitemOpen
  \bibfield  {author} {\bibinfo {author} {\bibfnamefont {A.}~\bibnamefont
  {Gali}}, \bibinfo {author} {\bibfnamefont {T.}~\bibnamefont {Simon}}, \ and\
  \bibinfo {author} {\bibfnamefont {J.~E.}\ \bibnamefont {Lowther}},\
  }\href@noop {} {\bibfield  {journal} {\bibinfo  {journal} {New Journal of
  Physics}\ }\textbf {\bibinfo {volume} {13}},\ \bibinfo {pages} {025016}
  (\bibinfo {year} {2011})}\BibitemShut {NoStop}%
\bibitem [{\citenamefont {van Oort}\ and\ \citenamefont
  {Glasbeek}(1989)}]{VAN1989}%
  \BibitemOpen
  \bibfield  {author} {\bibinfo {author} {\bibfnamefont {E.}~\bibnamefont {van
  Oort}}\ and\ \bibinfo {author} {\bibfnamefont {M.}~\bibnamefont {Glasbeek}},\
  }\href {\doibase 10.1103/PhysRevB.40.6509} {\bibfield  {journal} {\bibinfo
  {journal} {Phys. Rev. B}\ }\textbf {\bibinfo {volume} {40}},\ \bibinfo
  {pages} {6509} (\bibinfo {year} {1989})}\BibitemShut {NoStop}%
\bibitem [{\citenamefont {van Oort}\ and\ \citenamefont
  {Glasbeek}(1991)}]{VAN1991AMR}%
  \BibitemOpen
  \bibfield  {author} {\bibinfo {author} {\bibfnamefont {E.}~\bibnamefont {van
  Oort}}\ and\ \bibinfo {author} {\bibfnamefont {M.}~\bibnamefont {Glasbeek}},\
  }\href {http://dx.doi.org/10.1007/BF03166042} {\bibfield  {journal} {\bibinfo
   {journal} {Applied Magnetic Resonance}\ }\textbf {\bibinfo {volume} {2}},\
  \bibinfo {pages} {291} (\bibinfo {year} {1991})}\BibitemShut {NoStop}%
\bibitem [{\citenamefont {Armstrong}\ \emph {et~al.}(2010)\citenamefont
  {Armstrong}, \citenamefont {Rogers}, \citenamefont {McMurtrie},\ and\
  \citenamefont {Manson}}]{ARM2010}%
  \BibitemOpen
  \bibfield  {author} {\bibinfo {author} {\bibfnamefont {S.}~\bibnamefont
  {Armstrong}}, \bibinfo {author} {\bibfnamefont {L.~J.}\ \bibnamefont
  {Rogers}}, \bibinfo {author} {\bibfnamefont {R.~L.}\ \bibnamefont
  {McMurtrie}}, \ and\ \bibinfo {author} {\bibfnamefont {N.~B.}\ \bibnamefont
  {Manson}},\ }\href@noop {} {\bibfield  {journal} {\bibinfo  {journal}
  {Physics Procedia}\ }\textbf {\bibinfo {volume} {3}},\ \bibinfo {pages} {1569
  } (\bibinfo {year} {2010})}\BibitemShut {NoStop}%
\bibitem [{\citenamefont {Holliday}\ \emph {et~al.}(1989)\citenamefont
  {Holliday}, \citenamefont {Manson}, \citenamefont {Glasbeek},\ and\
  \citenamefont {van Oort}}]{HOL1989}%
  \BibitemOpen
  \bibfield  {author} {\bibinfo {author} {\bibfnamefont {K.}~\bibnamefont
  {Holliday}}, \bibinfo {author} {\bibfnamefont {N.~B.}\ \bibnamefont
  {Manson}}, \bibinfo {author} {\bibfnamefont {M.}~\bibnamefont {Glasbeek}}, \
  and\ \bibinfo {author} {\bibfnamefont {E.}~\bibnamefont {van Oort}},\ }\href
  {http://stacks.iop.org/0953-8984/1/i=39/a=021} {\bibfield  {journal}
  {\bibinfo  {journal} {Journal of Physics: Condensed Matter}\ }\textbf
  {\bibinfo {volume} {1}},\ \bibinfo {pages} {7093} (\bibinfo {year}
  {1989})}\BibitemShut {NoStop}%
\bibitem [{Note1()}]{Note1}%
  \BibitemOpen
  \bibinfo {note} {Hyperfine structure associated with $^{14}$N is not resolved
  in the present experiments.}\BibitemShut {Stop}%
\bibitem [{\citenamefont {Mizuochi}\ \emph {et~al.}(2009)\citenamefont
  {Mizuochi}, \citenamefont {Neumann}, \citenamefont {Rempp}, \citenamefont
  {Beck}, \citenamefont {Jacques}, \citenamefont {Siyushev}, \citenamefont
  {Nakamura}, \citenamefont {Twitchen}, \citenamefont {Watanabe}, \citenamefont
  {Yamasaki}, \citenamefont {Jelezko},\ and\ \citenamefont
  {Wrachtrup}}]{MIZ2009}%
  \BibitemOpen
  \bibfield  {author} {\bibinfo {author} {\bibfnamefont {N.}~\bibnamefont
  {Mizuochi}}, \bibinfo {author} {\bibfnamefont {P.}~\bibnamefont {Neumann}},
  \bibinfo {author} {\bibfnamefont {F.}~\bibnamefont {Rempp}}, \bibinfo
  {author} {\bibfnamefont {J.}~\bibnamefont {Beck}}, \bibinfo {author}
  {\bibfnamefont {V.}~\bibnamefont {Jacques}}, \bibinfo {author} {\bibfnamefont
  {P.}~\bibnamefont {Siyushev}}, \bibinfo {author} {\bibfnamefont
  {K.}~\bibnamefont {Nakamura}}, \bibinfo {author} {\bibfnamefont {D.~J.}\
  \bibnamefont {Twitchen}}, \bibinfo {author} {\bibfnamefont {H.}~\bibnamefont
  {Watanabe}}, \bibinfo {author} {\bibfnamefont {S.}~\bibnamefont {Yamasaki}},
  \bibinfo {author} {\bibfnamefont {F.}~\bibnamefont {Jelezko}}, \ and\
  \bibinfo {author} {\bibfnamefont {J.}~\bibnamefont {Wrachtrup}},\ }\href@noop
  {} {\bibfield  {journal} {\bibinfo  {journal} {Physical Review B}\ }\textbf
  {\bibinfo {volume} {80}},\ \bibinfo {pages} {041201} (\bibinfo {year}
  {2009})}\BibitemShut {NoStop}%
\bibitem [{\citenamefont {Kubo}\ \emph {et~al.}(2011)\citenamefont {Kubo},
  \citenamefont {Grezes}, \citenamefont {Dewes}, \citenamefont {Umeda},
  \citenamefont {Isoya}, \citenamefont {Sumiya}, \citenamefont {Morishita},
  \citenamefont {Abe}, \citenamefont {Onoda}, \citenamefont {Ohshima},
  \citenamefont {Jacques}, \citenamefont {Dr\'eau}, \citenamefont {Roch},
  \citenamefont {Diniz}, \citenamefont {Auffeves}, \citenamefont {Vion},
  \citenamefont {Esteve},\ and\ \citenamefont {Bertet}}]{KUB2011}%
  \BibitemOpen
  \bibfield  {author} {\bibinfo {author} {\bibfnamefont {Y.}~\bibnamefont
  {Kubo}}, \bibinfo {author} {\bibfnamefont {C.}~\bibnamefont {Grezes}},
  \bibinfo {author} {\bibfnamefont {A.}~\bibnamefont {Dewes}}, \bibinfo
  {author} {\bibfnamefont {T.}~\bibnamefont {Umeda}}, \bibinfo {author}
  {\bibfnamefont {J.}~\bibnamefont {Isoya}}, \bibinfo {author} {\bibfnamefont
  {H.}~\bibnamefont {Sumiya}}, \bibinfo {author} {\bibfnamefont
  {N.}~\bibnamefont {Morishita}}, \bibinfo {author} {\bibfnamefont
  {H.}~\bibnamefont {Abe}}, \bibinfo {author} {\bibfnamefont {S.}~\bibnamefont
  {Onoda}}, \bibinfo {author} {\bibfnamefont {T.}~\bibnamefont {Ohshima}},
  \bibinfo {author} {\bibfnamefont {V.}~\bibnamefont {Jacques}}, \bibinfo
  {author} {\bibfnamefont {A.}~\bibnamefont {Dr\'eau}}, \bibinfo {author}
  {\bibfnamefont {J.-F.}\ \bibnamefont {Roch}}, \bibinfo {author}
  {\bibfnamefont {I.}~\bibnamefont {Diniz}}, \bibinfo {author} {\bibfnamefont
  {A.}~\bibnamefont {Auffeves}}, \bibinfo {author} {\bibfnamefont
  {D.}~\bibnamefont {Vion}}, \bibinfo {author} {\bibfnamefont {D.}~\bibnamefont
  {Esteve}}, \ and\ \bibinfo {author} {\bibfnamefont {P.}~\bibnamefont
  {Bertet}},\ }\href {\doibase 10.1103/PhysRevLett.107.220501} {\bibfield
  {journal} {\bibinfo  {journal} {Phys. Rev. Lett.}\ }\textbf {\bibinfo
  {volume} {107}},\ \bibinfo {pages} {220501} (\bibinfo {year}
  {2011})}\BibitemShut {NoStop}%
\bibitem [{\citenamefont {Kubo}\ \emph {et~al.}(2010)\citenamefont {Kubo},
  \citenamefont {Ong}, \citenamefont {Bertet}, \citenamefont {Vion},
  \citenamefont {Jacques}, \citenamefont {Zheng}, \citenamefont {Dr\'eau},
  \citenamefont {Roch}, \citenamefont {Auffeves}, \citenamefont {Jelezko},
  \citenamefont {Wrachtrup}, \citenamefont {Barthe}, \citenamefont {Bergonzo},\
  and\ \citenamefont {Esteve}}]{KUB2010}%
  \BibitemOpen
  \bibfield  {author} {\bibinfo {author} {\bibfnamefont {Y.}~\bibnamefont
  {Kubo}}, \bibinfo {author} {\bibfnamefont {F.~R.}\ \bibnamefont {Ong}},
  \bibinfo {author} {\bibfnamefont {P.}~\bibnamefont {Bertet}}, \bibinfo
  {author} {\bibfnamefont {D.}~\bibnamefont {Vion}}, \bibinfo {author}
  {\bibfnamefont {V.}~\bibnamefont {Jacques}}, \bibinfo {author} {\bibfnamefont
  {D.}~\bibnamefont {Zheng}}, \bibinfo {author} {\bibfnamefont
  {A.}~\bibnamefont {Dr\'eau}}, \bibinfo {author} {\bibfnamefont {J.-F.}\
  \bibnamefont {Roch}}, \bibinfo {author} {\bibfnamefont {A.}~\bibnamefont
  {Auffeves}}, \bibinfo {author} {\bibfnamefont {F.}~\bibnamefont {Jelezko}},
  \bibinfo {author} {\bibfnamefont {J.}~\bibnamefont {Wrachtrup}}, \bibinfo
  {author} {\bibfnamefont {M.~F.}\ \bibnamefont {Barthe}}, \bibinfo {author}
  {\bibfnamefont {P.}~\bibnamefont {Bergonzo}}, \ and\ \bibinfo {author}
  {\bibfnamefont {D.}~\bibnamefont {Esteve}},\ }\href {\doibase
  10.1103/PhysRevLett.105.140502} {\bibfield  {journal} {\bibinfo  {journal}
  {Phys. Rev. Lett.}\ }\textbf {\bibinfo {volume} {105}},\ \bibinfo {pages}
  {140502} (\bibinfo {year} {2010})}\BibitemShut {NoStop}%
\bibitem [{\citenamefont {Ams\"uss}\ \emph {et~al.}(2011)\citenamefont
  {Ams\"uss}, \citenamefont {Koller}, \citenamefont {N\"obauer}, \citenamefont
  {Putz}, \citenamefont {Rotter}, \citenamefont {Sandner}, \citenamefont
  {Schneider}, \citenamefont {Schramb\"ock}, \citenamefont {Steinhauser},
  \citenamefont {Ritsch}, \citenamefont {Schmiedmayer},\ and\ \citenamefont
  {Majer}}]{AMS2011}%
  \BibitemOpen
  \bibfield  {author} {\bibinfo {author} {\bibfnamefont {R.}~\bibnamefont
  {Ams\"uss}}, \bibinfo {author} {\bibfnamefont {C.}~\bibnamefont {Koller}},
  \bibinfo {author} {\bibfnamefont {T.}~\bibnamefont {N\"obauer}}, \bibinfo
  {author} {\bibfnamefont {S.}~\bibnamefont {Putz}}, \bibinfo {author}
  {\bibfnamefont {S.}~\bibnamefont {Rotter}}, \bibinfo {author} {\bibfnamefont
  {K.}~\bibnamefont {Sandner}}, \bibinfo {author} {\bibfnamefont
  {S.}~\bibnamefont {Schneider}}, \bibinfo {author} {\bibfnamefont
  {M.}~\bibnamefont {Schramb\"ock}}, \bibinfo {author} {\bibfnamefont
  {G.}~\bibnamefont {Steinhauser}}, \bibinfo {author} {\bibfnamefont
  {H.}~\bibnamefont {Ritsch}}, \bibinfo {author} {\bibfnamefont
  {J.}~\bibnamefont {Schmiedmayer}}, \ and\ \bibinfo {author} {\bibfnamefont
  {J.}~\bibnamefont {Majer}},\ }\href {\doibase 10.1103/PhysRevLett.107.060502}
  {\bibfield  {journal} {\bibinfo  {journal} {Phys. Rev. Lett.}\ }\textbf
  {\bibinfo {volume} {107}},\ \bibinfo {pages} {060502} (\bibinfo {year}
  {2011})}\BibitemShut {NoStop}%
\bibitem [{\citenamefont {Zhu}\ \emph {et~al.}(2011)\citenamefont {Zhu},
  \citenamefont {Saito}, \citenamefont {Kemp}, \citenamefont {Kakuyanagi},
  \citenamefont {Karimoto}, \citenamefont {Nakano}, \citenamefont {Munro},
  \citenamefont {Tokura}, \citenamefont {Everitt}, \citenamefont {Nemoto},
  \citenamefont {Kasu}, \citenamefont {Mizuochi},\ and\ \citenamefont
  {Semba}}]{ZHU2011}%
  \BibitemOpen
  \bibfield  {author} {\bibinfo {author} {\bibfnamefont {X.}~\bibnamefont
  {Zhu}}, \bibinfo {author} {\bibfnamefont {S.}~\bibnamefont {Saito}}, \bibinfo
  {author} {\bibfnamefont {A.}~\bibnamefont {Kemp}}, \bibinfo {author}
  {\bibfnamefont {K.}~\bibnamefont {Kakuyanagi}}, \bibinfo {author}
  {\bibfnamefont {S.-i.}\ \bibnamefont {Karimoto}}, \bibinfo {author}
  {\bibfnamefont {H.}~\bibnamefont {Nakano}}, \bibinfo {author} {\bibfnamefont
  {W.~J.}\ \bibnamefont {Munro}}, \bibinfo {author} {\bibfnamefont
  {Y.}~\bibnamefont {Tokura}}, \bibinfo {author} {\bibfnamefont {M.~S.}\
  \bibnamefont {Everitt}}, \bibinfo {author} {\bibfnamefont {K.}~\bibnamefont
  {Nemoto}}, \bibinfo {author} {\bibfnamefont {M.}~\bibnamefont {Kasu}},
  \bibinfo {author} {\bibfnamefont {N.}~\bibnamefont {Mizuochi}}, \ and\
  \bibinfo {author} {\bibfnamefont {K.}~\bibnamefont {Semba}},\ }\href@noop {}
  {\bibfield  {journal} {\bibinfo  {journal} {Nature}\ }\textbf {\bibinfo
  {volume} {478}},\ \bibinfo {pages} {221} (\bibinfo {year}
  {2011})}\BibitemShut {NoStop}%
\end{thebibliography}
\end{document}


\title{Supplemental material for: "Temperature and magnetic field dependent longitudinal spin relaxation in nitrogen-vacancy ensembles in diamond"}

\author{A.~Jarmola}
 \email{jarmola@berkeley.edu}
    \affiliation{
     Department of Physics, University of California,
     Berkeley, California 94720, USA
     }
    \affiliation{
    Laser Centre, The University of Latvia, Rainis Boulevard 19, 1586 Riga, Latvia
    }
 \author{V.~M.~Acosta}
    \affiliation{
     Department of Physics, University of California,
     Berkeley, California 94720, USA
    }
     \affiliation{
     Hewlett-Packard Laboratories, 1501 Page Mill Road, Palo Alto, California 94304, USA
    }

 \author{K.~Jensen}
    \affiliation{
     Department of Physics, University of California,
     Berkeley, California 94720, USA
    }

\author{S.~Chemerisov}
    \affiliation{
     Argonne National Laboratory, Argonne, Illinois 60439, USA
    }

 \author{D.~Budker}
 \email{budker@berkeley.edu}
    \affiliation{
     Department of Physics, University of California,
     Berkeley, California 94720, USA
    }
    \affiliation{Nuclear Science Division, Lawrence Berkeley National Laboratory,
     Berkeley, California 94720, USA
    }

\maketitle

\section{Temperature-dependence fits for individual samples}

\begin{figure}
\centering
    \includegraphics[width=1.0\columnwidth]{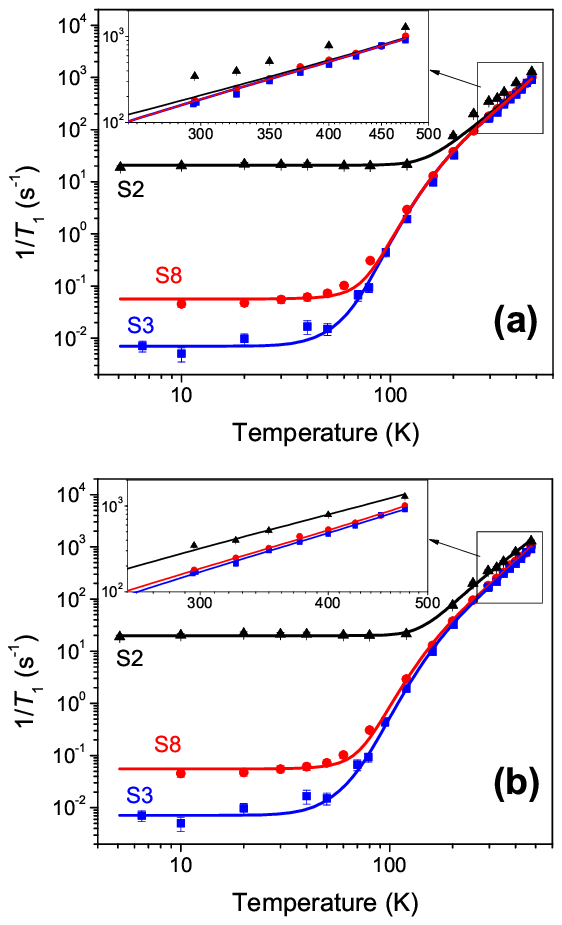}
    \caption{\label{fig:fit} $1/T_{1}$ as a function of temperature for S2, S3, and S8 samples. Symbols represent experimental data while solid lines are fits to the model in Eq. (1) where (a) all parameters except $A_{1}$(S) are common for all samples and (b) all parameters are variable.}
\end{figure}

In the main text, we presented a global fit of the temperature dependence of $1/T_1$ for the three samples investigated in this work. The global fit (Fig.\ 2(a) in the main text and Fig. \ref{fig:fit}(a) here) describes well the $1/T_1$ variation over some five orders of magnitude. However, at the next ("factor-of-two") level of detail, the experimentally significant deviations for the high-temperature part of the dependence for the highest-NV-density S2 sample, clearly seen in the insets of Fig. \ref{fig:fit}, become important. In order to understand these deviations, we performed fits of the temperature dependence given by Eq.\ (1) in the main text individually for each of the three samples studied in this work (Fig. \ref{fig:fit}(b)). The values of the fit parameters are collected in Table \ref{tab:fit_values}. The fit results indicate that [within the model of Eq.\ (1)] there is an additional contribution to the relaxation appears to be present in the highest-NV-density sample S2 at temperatures $\lesssim$ 100 K.

\begin{figure}
\centering
    \includegraphics[width=1.0\columnwidth]{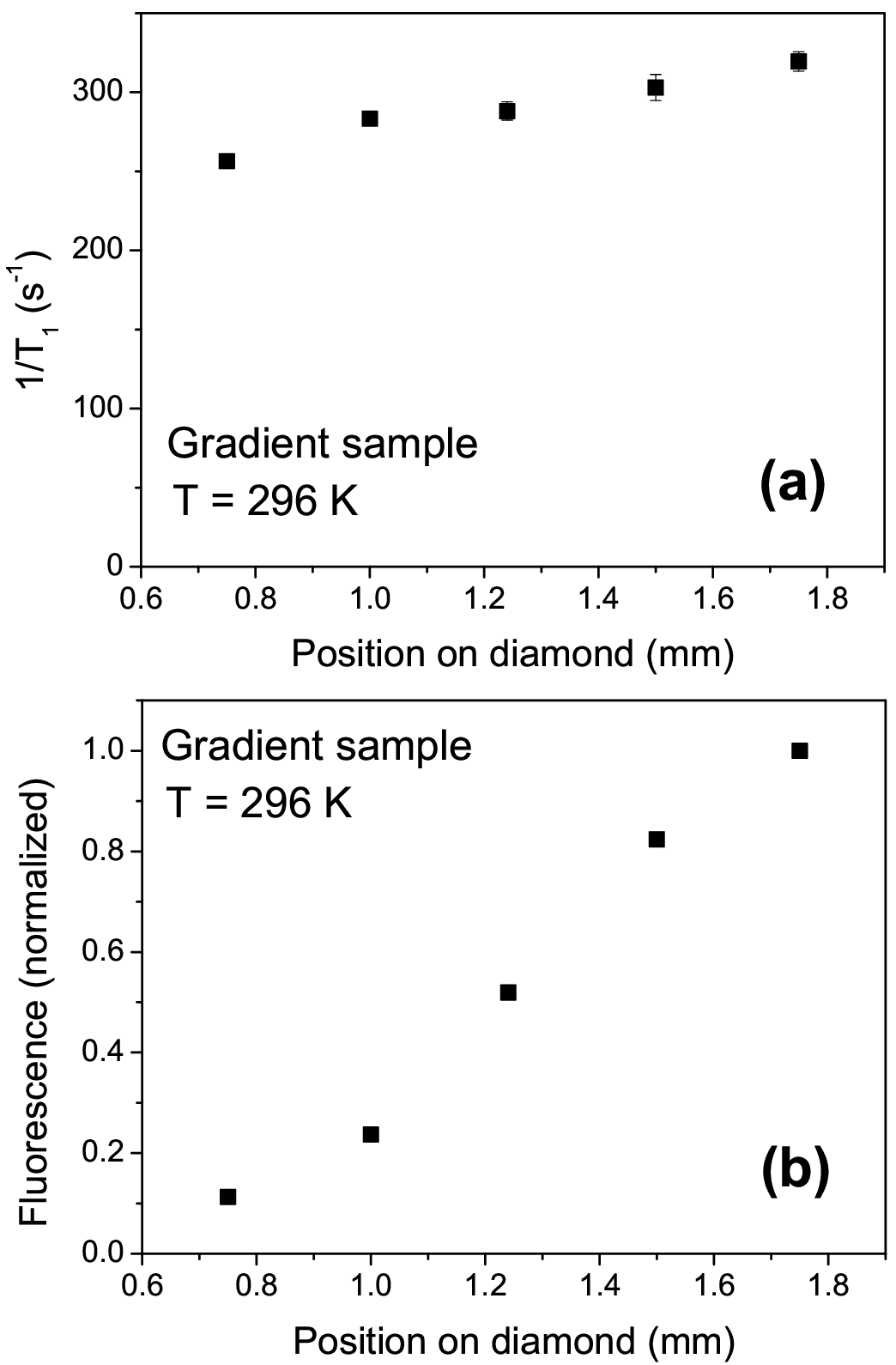}
    \caption{\label{fig:B7_296K} (a) $1/T_{1}$ and (b) fluorescence intensity at 296 K versus position on the sample with a spatially-varying electron irradiation dose.}
\end{figure}

A detailed investigation into the nature of this additional relaxation awaits further work; however, as an initial step in that direction, we have measured $1/T_1$ at 296 K for different locations on the  gradient sample mentioned in the main text at magnetic field, B $\approx$ 25 G applied parallel to the [100] crystallographic orientation (see Fig.\ 2(c)). The results are shown in Fig. \ref{fig:B7_296K}. The highest-NV-density part of the sample is similar in terms of its properties to the S2 sample. There appears to be a contribution to $1/T_1$ that is correlated with NV density, which one would not expect from Eq.\ (1) under the assumption that the coefficients in the second and third term are intrinsic to diamond.  We note that the magnetic-field dependence of relaxation shown in Fig. 3(a) in the main text also points towards a cross-relaxation contribution at room temperature exceeding what one would have expected from our simplified global picture.

\begin{table}[ht]
\caption{\label{tab:fit_values} Fit parameter values from fits of the Eq. (1) to the longitudinal spin relaxation data in Fig. \ref{fig:fit}. }
\centering
    \begin{ruledtabular}
    \begin{tabular}{c c c c c}
      \ Sample & $A_{1}$ (s$^{-1}$) & $A_{2}$ (s$^{-1}$) & $\Delta$ (meV) & $A_{3}$ (K$^{-5}$s$^{-1}$) \\
      \colrule
      & & Global fit & & \\
      \hline
      S2 & 21(3) &   &   &   \\
      S8 & 0.06(1) & $2.1(6) \times 10^{-3}$ & $73(4)$ & $2.2(5) \times 10^{-11}$ \\
      S3 & 0.007(4) &   &  & \\
      \hline
      & & Individual fits & & \\
      \hline
      S2 & 20(3) & $4.7(2.4) \times 10^{-3}$ & $77(6)$ & $2.1(8) \times 10^{-11}$ \\
      S8 & 0.06(1)& $1.7(6) \times 10^{-3}$ & $69(4)$ & $2.5(5) \times 10^{-11}$ \\
      S3 & 0.007(3) & $2.1(4) \times 10^{-3}$ & $76(4)$ & $2.2(3) \times 10^{-11}$ \\
    \end{tabular}
    \end{ruledtabular}
\end{table}

\section{$T_1$ comparison with other systems}

In view of the importance of long-lived electron-spin systems in applications (which, in addition to magnetometry, also include quantum information science, where the electron-spin $T_1$ is a limiting factor \cite {DUT2007, NEU2008}, in Table \ref{tab:T1_comp}, we list representative values of $T_1$ for both cryogenic and room temperatures. The Table clearly illustrates the advantages of the nitrogen vacancy centers, especially, at room temperature.

\begin{table}[ht]
\caption{\label{tab:T1_comp} Comparison of approximate longitudinal spin relaxation time values for different defects in solids which are of high interest for spin-based information processing devices }
\centering
    \begin{ruledtabular}
    \begin{tabular}{c c c c c}
      \ Defect type & $T$ (K) & B (G) & $T_{1}$ (s) & Reference \\
      \colrule
      NV in diamond & 300 & 25 & $6 \times 10^{-3}$ & This work \\
      NV in diamond & 10 & 25 & $2 \times 10^{2}$ & This work \\
      Defects in 4H-SiC & 300 & 0 & $2 \times 10^{-4}$ & \cite{KOE2011} \\
      Ga(In)As quantum dot & 1 & $4 \times 10^{4}$ & $2 \times 10^{-2}$ & \cite{KRO2004} \\
      Si/SiO$_{2}$ quantum dot & 0.05 & $2 \times 10^{4}$ & $4 \times 10^{-2}$ & \cite{XIA2010} \\
      Si/SiGe quantum dot & 0.05 & $1.5 \times 10^{4}$ & $2 \times 10^{-1}$ & \cite{HAY2009} \\
      P donors in Si & 10 & $3.5 \times 10^{3}$ & $1 \times 10^{-3}$ & \cite{TYR2012} \\
      P donors in Si & 1.8 & $3.5 \times 10^{3}$ & $2 \times 10^{3}$ & \cite{TYR2012} \\

    \end{tabular}
    \end{ruledtabular}
\end{table}